\documentclass[11pt,twoside,a4paper]{article}
\usepackage{graphics}
\usepackage[pdftex]{graphicx}
\usepackage[pdftex]{hyperref}
\usepackage{listings}
\usepackage{cite}
\usepackage{titlesec}

\usepackage{wrapfig}
\usepackage{caption}
\usepackage{subcaption}
\usepackage{sidecap}
\usepackage{url}

\begin{document}

\title{Zgoubi \\ A startup guide for the complete beginner}
\author{Annette Pressman and Kai Hock}
\date{\today}

\maketitle

\begin{abstract}
Zgoubi is a code which can be used to model accelerators and beam lines, comprised of magnetic and electrostatic elements. It has been extensively developed since the mid-1980s to include circular accelerators and related beam physics. It has been made freely available by its author on a code development site, including a Users' Guide, a data treatment/graphic interfacing tool, and many examples\cite{sourceforge}.

\end{abstract}

\newpage

\tableofcontents

\section{Getting started}

\subsection{Introduction to Zgoubi}

Zgoubi began in the early 1970’s as a code to be used for ray-tracing in large spectrometers for nuclear physics. Since then Zgoubi has been adapted such that it can be used to design beam optics and beam lines, and to simulate nonlinear dynamics. From the mid-1980's on, it has been developed and used extensively for the simulation of circular accelerators and storage rings, and related beam physics as spin dynamics, synchrotron radiation, short-lived beams, etc. It has been made freely available by its author on a code development site,  including a Users' Guide, a data treatment/graphic interface program, zpop, and many examples \cite{sourceforge, userguide, cern, examples}.

The code is written in FORTRAN but users have developed additional interfaces for data handling and visualisation. In this guide reference will be made to MATLAB \cite{matlab} scripts written for this purpose. 

	In the Zgoubi code, particle trajectories in electromagnetic fields are calculated through numerical integration of the Lorentz equations. As a result of the efficiency of the integration (through a Taylor-series based expansion) and the nature of the programming language used the computation time is very fast. The main program array is a linear description of the Zgoubi input file, named zgoubi.dat. This file contains a linear definition of the lattice and beam parameters, and is written for the specific simulation required.

\subsection{Installation of zgoubi}
	Zgoubi can be run on Windows or UNIX systems. However the post-processing program, zpop, requires an xterm window so is better suited to a UNIX system. On most Windows systems it is possible to run a UNIX system such as Scientific Linux (\url{https://www.scientificlinux.org/}) using a product such as VirtualBox (\url{https://www.virtualbox.org/}), both of which are freely available. During the installation of Scientific Linux it is possible to select the developer options which should include the required FORTRAN compilers needed for Zgoubi.
	
	The installation of Zgoubi on Windows and UNIX systems is covered in the following section. It is also possible to install Zgoubi through installation of PyZgoubi (see section \ref{sec:pyz})\cite{sam}. It is recommended that the new user gains some experience with Zgoubi to gain some understanding of how the code works before expanding to Zgoubi interfaces such as PyZgoubi. 
	
	\subsubsection{Required Software}
	\label{sec:req}
	
	To use Zgoubi you will need a FORTRAN compiler on your machine. Many UNIX systems will already have a compiler installed. On a Windows machine MinGW can be used \cite{min}.
	
	
	To install MinGW;
	
	\begin{enumerate}
	\item Visit \url{http://www.mingw.org/wiki/Getting_Started}.
	 \item Click on `mingw-get-inst', download and run the latest .exe version (figure \ref{min1}) to install to a directory of your choice, for example C:\textbackslash{MinGW}. Ensure to select the FORTRAN option during the install process (figure \ref{min2}). Further information about this install is available on the `Getting Started' webpage above.
\begin{figure}[!h]
\centering
	 \begin{subfigure}[b]{\textwidth}
                 \centering
                 \includegraphics[width=0.9\textwidth]{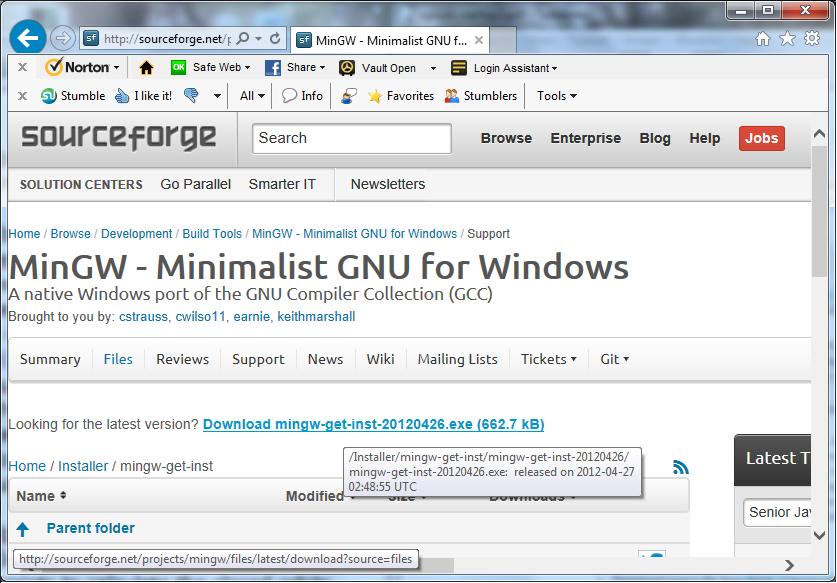}
                 \caption{The MinGW download file.}
                 \label{min1}
         \end{subfigure}
         ~\\
          \begin{subfigure}[b]{\textwidth}
                 \centering
                 \includegraphics[width=0.6\textwidth]{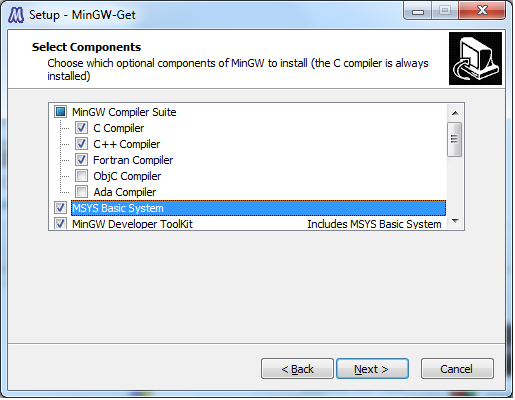}
                 \caption{Selecting MinGW install options.}
                 \label{min2}
         \end{subfigure}
         \caption{Installing MinGW.}
         \end{figure}
	 \item 	In the subfolder 'MinGW\textbackslash{bin}' rename the file 'mingw32-make.exe' to 'make.exe' (figure \ref{min3}).
	 	 \item Set the environmental variable with the address of the MinGW\textbackslash{bin} directory (see section \ref{sec:env}).
	 \begin{figure}
	 \centering
	 \includegraphics[width=0.7\textwidth]{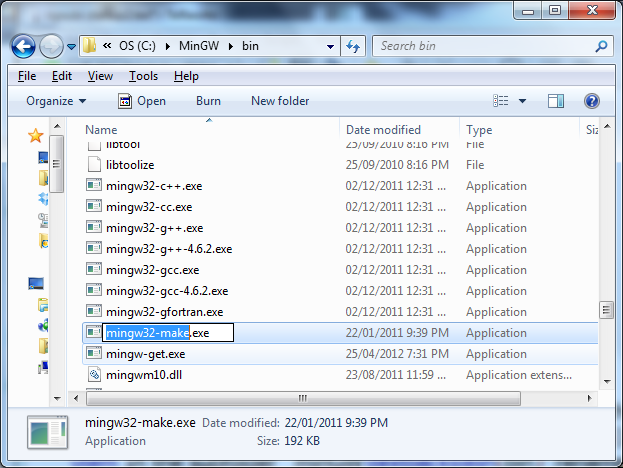}
	 \caption{Renaming to 'make.exe'.}
	 \label{min3}
	 \end{figure}
	\end{enumerate}

	To install Zgoubi;
	
	\begin{itemize}
	\item Windows
		
	\begin{enumerate}
	\item Download Zgoubi from source, e.g. \url{http://sourceforge.net/projects/zgoubi/} and unpack to a directory of your choosing, e.g. C:\textbackslash{zgoubi-5.1.0}. Version 5.1.0 is stable but newer versions are available.
	\item To use Zgoubi in Windows; in the main directory of the unpacked folder and every subfolder edit the 'Makefile' to replace every occurrence of ';' with '\&' (figure \ref{z2}): in UNIX the semicolon can be used separate two lines of command, but in Windows the ampersand performs this function. 
	\begin{figure}[h!]
	\centering
	\begin{subfigure}[b]{\textwidth}
	\centering
	\includegraphics[width=0.7\textwidth]{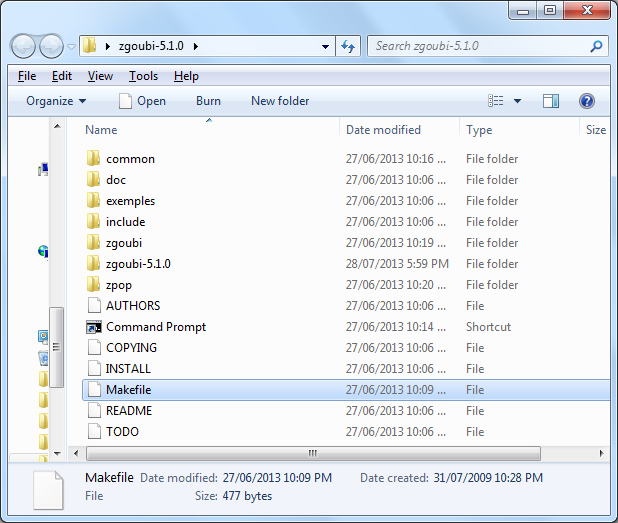}
	\caption{The zgoubi-5.1.0 contents.}
	\label{z1}
	\end{subfigure}
	~\\
	\begin{subfigure}[b]{\textwidth}
	\centering
	\includegraphics[width=0.7\textwidth]{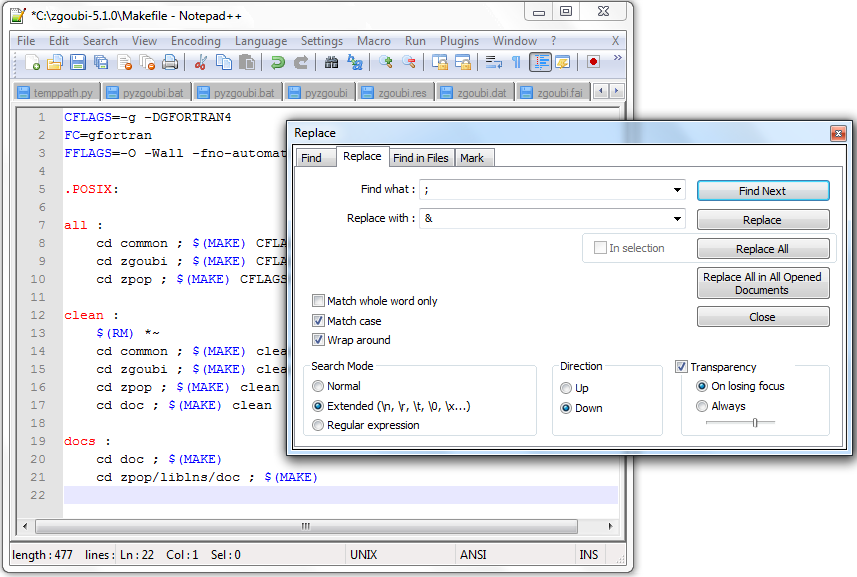}
	\caption{Editing 'Makefile'.}
	\label{z2}
	\end{subfigure}
	\caption{Editing the Zgoubi 'Makefile' to work in Windows.}
	\end{figure}

	\item Set the environmental variable with the address of the file 'zgoubi.exe'  (see section \ref{sec:env}).
	\item Open a command prompt (cmd.exe) in the Zgoubi folder and type the command \textbf{make}. This should trigger an install of the program (see figure \ref{make}). In Windows the install is likely to complete with error messages for the install of 'zpop'. This graphics package needs to be run in an xterm window, which requires an X11 environment. This is not a trivial process for most Windows systems so it will not be covered here.
	\begin{figure}[h!]
	\centering
	\includegraphics[width=0.7\textwidth]{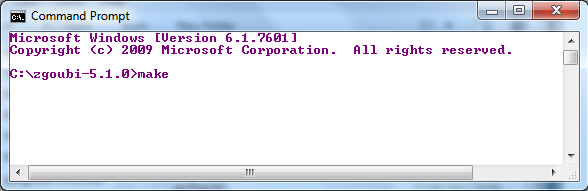}
	\caption{Entering the command 'make' on the Windows command line.}
	\label{make}
	\end{figure}
		\end{enumerate}
		
	\item UNIX
	\begin{enumerate}
		\item If you are unsure if there is a FORTRAN compiler on your system entering the command \textbf{apropos fortran} will show a list of the commands available and their use (figure \ref{apropos}). If there is no installer available it can be obtained using the command \textbf{sudo apt-get install gcc-gfortran}, or through the GUI software installer.
		\begin{figure}[h!]
		\centering
		\includegraphics[width=0.7\textwidth]{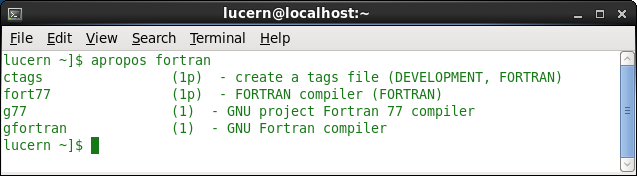}
		\caption{Expected response from the \textbf{apropos} command if FORTRAN compilers are installed.}
		\label{apropos}
		\end{figure}
		\item Download source file from \url{http://sourceforge.net/projects/zgoubi/} (e.g. zgoubi-5.1.0.tar.bz2).
		\item Unpack using the command \textbf{tar -xjvf zgoubi-5.1.0.tar.bz2} (figure \ref{zu1}).
		\begin{figure}[h!]
		\centering
		\includegraphics[width=0.7\textwidth]{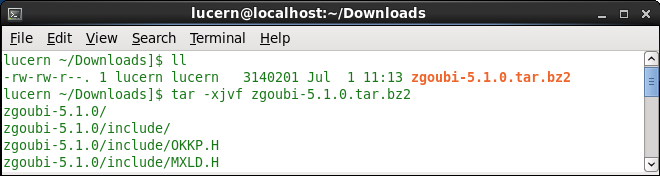}
		\caption{Unpacking Zgoubi download in Scientific Linux.}
		\label{zu1}
		\end{figure}
		\item 	Move the unpacked directory 'zgoubi-5.1.0' to a directory of your choice, e.g. usr/local (moved to folder \url{~}/sw in figure \ref{zu3}).
		\item Within the directory 'zgoubi-5.1.0' enter the command \textbf{make} (figure \ref{zu3}).
		\begin{figure}
		\centering
		\includegraphics[width=0.7\textwidth]{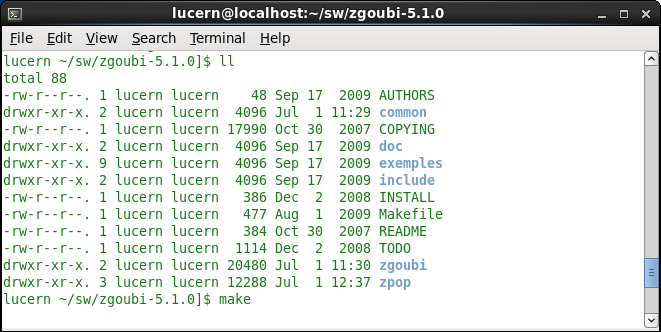}
		\caption{Entering the \textbf{make} command in Scientific Linux.}
		\label{zu3}
		\end{figure}
		\item Add the location of the folder `zgoubi-5.1.0/zgoubi' to the Path - see section \ref{sec:env}.
	\end{enumerate}
	
	\end{itemize}
	
\subsubsection{Setting Environmental Variables}
\label{sec:env}
	Both MinGW and Zgoubi will need adding to the `Path' variable.
In order to do this, follow the instructions for the operating system used.
	
	\begin{itemize}
	\item	Windows 
	\begin{enumerate}
		\item From the Start menu right-click on `My Computer' or `Computer' and select `Properties', and 'Advanced System Settings'.
		\item Under the `Advanced' tab select `Environmental Variables' (figure \ref{env}).
\begin{figure}[h!]
\centering
\includegraphics[width=\textwidth]{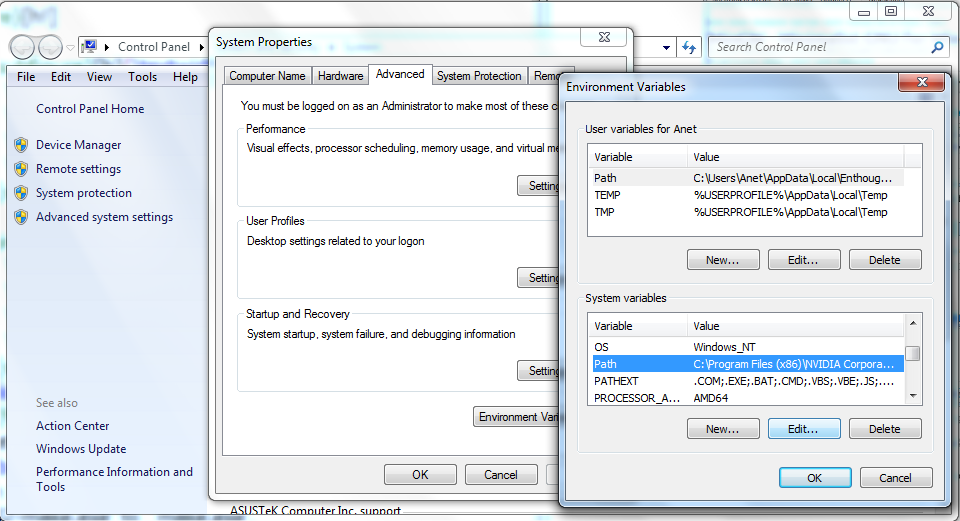}
\caption{Setting the Path in Windows 7.}
\label{env}
\end{figure}
		\item Here there are two boxes, the upper specifying user settings and the lower specifying settings for all users. Be careful editing these settings as a mistake can affect the correct operation of the computer.
		\item In the lower box highlight 'Path' and click 'Edit'.
		\item At the end of the existing entry add the address for the MinGW bin directory and the Zgoubi directory, e.g. for the base directories described in the install above the add to the path ';C:\textbackslash{MinGW}\textbackslash{bin}' and ';C:\textbackslash{zgoubi-5.1.0}' (figure \ref{path}).
		\begin{figure}[h!]
\centering
\includegraphics[width=0.6\textwidth]{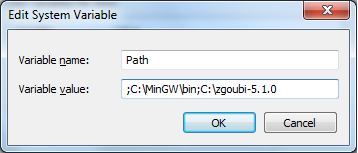}
\caption{Inputing the path addresses.}
\label{path}
\end{figure}
	\end{enumerate}

	\item UNIX
	\begin{enumerate}
		\item Edit the startup file and modify the PATH variable to include the location of the file /zgoubi-5.1.0/zgoubi as follows (e.g. if this file is in the directory usr/local/);
		\begin{itemize}
			\item bash shell
			\begin{enumerate}
				\item edit the `\url{~}/ .bashrc' file with the folowing;
				\item PATH="\$PATH":/usr/local/zgoubi-5.1.0/zgoubi
				\item export PATH
			\end{enumerate}
			\item C shell
			\begin{enumerate}
				\item edit the `\url{~}/ .cshrc' file with the following;
				\item set path="\$PATH":/usr/local/zgoubi-5.1.0/zgoubi
			\end{enumerate}	
		\end{itemize}
		\item Save and close the file.
		\item To use the graphics program `zpop' another line can be added to the file using the location of the file `/zgoubi-5.1.0/zpop' in the same manner as above (inserted before the `export PATH' line for the bash shell) as can be seen in figure \ref{bash2} below.
	\end{enumerate}
	
	\begin{figure}[h!]
	\centering
	\begin{subfigure}[b]{\textwidth}
	\centering
	\includegraphics[width=0.6\textwidth]{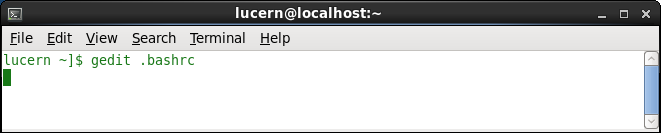}
	\caption{Opening the bash shell script in the home directory with text editor 'gedit'.}
	\label{bash1}
	\end{subfigure}
	~\\
	\begin{subfigure}[b]{\textwidth}
	\centering
	\includegraphics[width=0.6\textwidth]{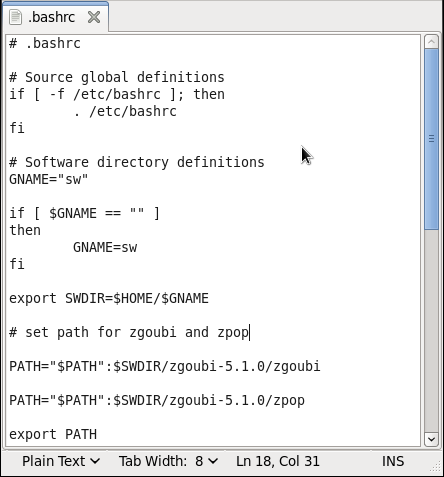}
	\caption{The completed shell script.}
	\label{bash2}
	\end{subfigure}
	\caption{Editing the shell script in Scientific Linux.}
	\end{figure}
	
	\end{itemize}
	
\subsubsection{Hints}
	Depending on your familiarity with Windows the following information may be unnecessary.
	
	\begin{itemize}
	\item	You will need a text editor to edit FORTRAN files.  `Notepad++' (\url{http://notepad-plus-plus.org/}) is freely available and an easy to use option for Windows.
	\item	In order to quickly open the command prompt in the correct folder on each use, paste a shortcut to the cmd.exe into the example folder, right click and under `Properties', in the `Start in' field copy and paste the address of the folder containing the zgoubi.dat file.
	\end{itemize}
	
\subsection{Running Zgoubi}

\begin{figure}[h!]
\centering
\includegraphics[width=0.6\textwidth]{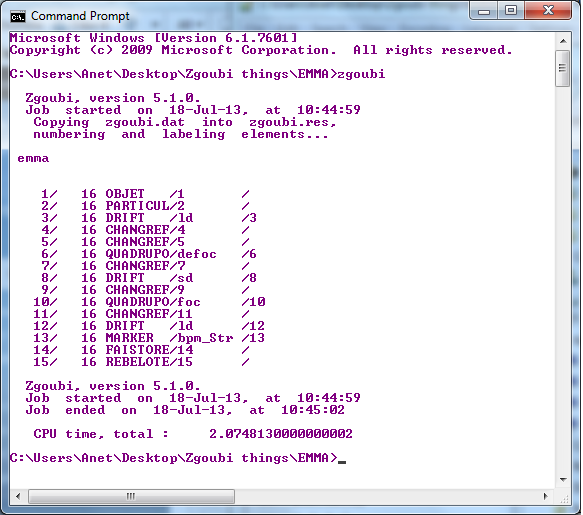}
\caption{Running Zgoubi in Windows.}
\label{emmarun}
\end{figure}

	In Windows the program is run from the command line (cmd.exe). In both Windows and Linux, at the command line within the folder containing the zgoubi.dat file to be used, type \textbf{zgoubi} and enter (see figure \ref{emmarun}). 
	
	A successful run will output a list of elements as specified in the .dat file and produce a results file, `zgoubi.res' alongside other files such as;
	\begin{itemize}
	\item zgoubi.fai - particle coordinates and spin information. This stores data at the end of each magnet - it is useful for analysing phase space.
	\item zgoubi.spn - spin coordinate information.
	\item zgoubi.plt - particle coordinates and fields experienced by the particles stepwise along the beamline. This stores data points within a magnetic field only, drift spaces don't store tracking data in this file.
	\item zgoubi.map - 2-D field map information.
	\end{itemize}

	The results output overview of Zgoubi is written to the  zgoubi.res file. The particle tracking and field data are stored to the other output files. The data can be visualised using the Zpop code for users with access to an xterm window. Otherwise extracting the required data can be done using other program interfaces. Dr Kai Hock has written MATLAB scripts to perform this function (see section \ref{sec:scripts}). There are also add-on scripts such as PyZgoubi which uses a python interface to input lattice data and analyse the results (see section \ref{sec:pyz}).
	
	Further information about the Zgoubi output files can be found in Part D of the Zgoubi User's Guide. 
	
\subsection{Running zpop}
	The Zgoubi code comes with a graphics package called zpop. It can only be run in an xterm window which can be accessed on most UNIX systems with the command \textbf{xterm}. An example of use can be seen in section \ref{sec:zpop}.

\subsection{How to use the Zgoubi User's Guide}
	The Zgoubi User's Guide is split into four sections. The first section, Part A - `Description of software contents', provides information about the physics processes considered, the procedures available and the mathematical methods used to perform the integrations. The section begins with a glossary of the keywords used in the program. The keywords are how the user defines the accelerator lattice in the input zgoubi.dat file. Further information about the keywords can be found in section 4 of Part A, with details about how to use the keywords to be found in section B.
	
	Part B - `Keywords and input data formatting' details all the keywords used in the program. To enter lattice information into the zgoubi.dat file a keyword is used, followed by further numerical or text data to define the use of the keyword. As a simple example a drift section is defined in zgoubi.dat with the keyword `DRIFT' with the length of the section (in cm). A few other examples will be discussed in section \ref{sec:define} of this guide. 
	
	In the introduction to Part B a description is given of how to input the information given for each keyword. An important thing to note is the units expected for each keyword. As an example, both radians and degrees are used at different occasions. When defining a lattice careful reference to the use of the keywords as defined in this section is strongly recommended. The allowed values for each keyword and the units to be used can be found in part B.
	
	In Part C - `Examples of input data files and output result files' several zgoubi.dat files are given as examples of different accelerator definitions. The resultant zgoubi.res files are also shown.
	
	The information regarding instillation and running of Zgoubi is contained within Part D - `Running Zgoubi and its post-processor/graphic interface zpop'. 
	
\subsection{PyZgoubi - a python interface to Zgoubi}
\label{sec:pyz}
	PyZgoubi is a freely available interface to Zgoubi written in python (\url{http://www.hep.manchester.ac.uk/u/sam/pyzgoubi/}). It has been designed to make reading and writing input data files easier, and takes advantage of python's utilities and analysis tools to interpret the output\cite{sam2}. It requires an installation of python with the packages of NumPy, SciPy and Matplotlib. It is also possible to install Zgoubi from the PyZgoubi installation if the OS has the GFortran compiler installed.

\subsubsection{Installing PyZgoubi on Windows}

	Python can be installed on Windows with the required packages and libraries, NumPy, SciPy, Matplotlib either as individual modules ( see \url{http://www.scipy.org/}) or through the single package options such as Anaconda (\url{http://continuum.io/downloads}) or Enthought Canopy (\url{https://www.enthought.com/products/epd/free/}). The 32 bit version is free to all users. Links to these products and more are available on the PyZgoubi homepage above. The environmental variables need be amended to include the executable 'python.exe' and also the library site packages, which may have an address such as '\verb+C:\Users\Username\AppData\Local\Enthought\Canopy32+ \verb+\User\Lib\site-packages+'.
	
	After installing a python package with the above libraries PyZgoubi can be downloaded from sourceforge via the homepage (\url{http://www.hep.manchester.ac.uk/u/sam/pyzgoubi/}). After downloading 	it can be unpacked  to a directory such as 'C:\textbackslash{pyzgoubi}' through a MinGW  (see \ref{sec:req} for MinGW information) interface opened in the directory containing the tar.gz file (for example 'filename.tar.gz') using the following command;

	\begin{lstlisting}	
	tar -zxvf filename.tar.gz
	\end{lstlisting}
	
	After unpacking PyZgoubi, through a command prompt open in the pyzgoubi folder enter the command \textbf{python setup.py install}, or to install to a specific destination (e.g. /python/install/) with the command \textbf{python setup.py install \texttt{-{}-}prefix=\url{~}/python/install/}. The installation dialogue will offer suggestions of amendments to be added to the environmental Path variable which should be updated accordingly, along with the address of the PyZgoubi executable (see section \ref{sec:env}). 
	
	Depending on the installation some small amendments may be necessary to run PyZgoubi. A batch file in the pyzgoubi folder (titled for example 'pyzgoubi.bat') can help pyzgoubi to locate the python.exe executable. An example is shown below for an Enthought installation of python.
	
\begin{lstlisting}
@echo off
python C:\Users\Username\AppData\Local\Enthought\Canopy32\User\Scripts\pyzgoubi %1
\end{lstlisting}
	
	In the python file 'core.py' within the 'run(...  )' loop a temporary directory is made with a prefix. This can cause problems in Windows which can be patched by changing the definition of 'tmpdir = tempfile.mdktemp(... )' in this loop to 'tmpdir = tempfile.mkdtemp()'. On some systems the 'settings.ini' file in the .pyzgoubi folder may need to be edited to ensure the path names are correct.
	
\subsubsection{Installing PyZgoubi on UNIX}
	The PyZgoubi homepage contains links and information about how to install the code. The required packages of SciPy, NumPy and Matplotlib can be installed through Anaconda (link from the homepage) or other sources. Anaconda will also suggest appropriate changes to the PATH variable or offer to adjust the shell script automatically. 
	
	After a successful installation of python with these libraries on a system with a g77 compiler it should now be possible to download and install PyZgoubi following the advice on the homepage and in the README included with the source code. The shell environment needs to be configured to include the paths for the python executable and paths for the python libraries (PYTHONPATH) as advised during the installation procedure.
	
\newpage
\section{Defining a lattice}
\label{sec:define}

	To define an accelerator lattice for Zgoubi the details need to be saved into a text file named `zgoubi.dat'. Keeping this input file for different accelerators in separate directories is essential to avoid accidentally overwriting  one with another. In the file each element of the accelerator is specified in turn.
	
\subsection{Coordinates}

\begin{figure}[h!]
 \begin{center}
 \includegraphics[width=0.8\textwidth]{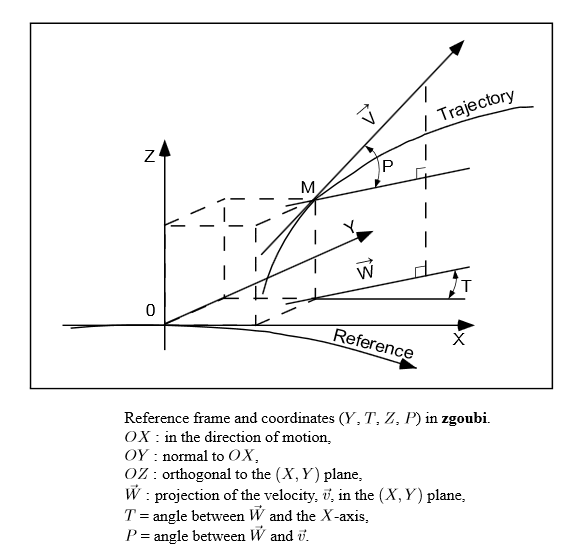}
\end{center}
\caption{The Zgoubi coordinate reference frame. Taken from \cite{userguide}}
\label{coord}
\end{figure}

 	 It is important to note that Zgoubi always works in local coordinates, not global coordinates. A right-handed system is used but the terminology may be slightly different to other commonly used systems, most notably in that the x axis is along the general direction of motion. The x axis is zeroed at the beginning of a component. In the figure below (see figure \ref{coord}) the origin is in the median plane on a reference curve which coincides with the optical axis of optical elements. 
 	 
\subsection{Keywords and labels}

	In the zgoubi.dat file a keyword is used (in capitals and in quotes) to input data about every step of the lattice. The keyword and the data following it is read into the Zgoubi program. Each keyword can have two 'labels' on the same line.
	
\begin{lstlisting}
'KEYWORD' label1 label2
\end{lstlisting}

 Although this label can be free text and used as an aide memoir, it can also on occasion be used to produce a result. For the keyword 'MARKER', if the second label is \textbf{.plt} the current coordinates will be stored to the file named 'zgoubi.plt'. This file can be used with Zpop to produce plots. As another example, if given the name of a label the keyword 'FAISTORE' will create a '.fai' file for storage of particle data at any element where the label is used. An example is given below. Here the output file will be called 'zgoubi.fai' and will store data at the exit of any element labelled \textbf{label1}.
 
 \begin{lstlisting}
 'FAISTORE'
 zgoubi.fai label1
 \end{lstlisting}
 
	The details about the available keywords and how to use them can be found in Part B of the Zgoubi User's Manual. Further information and examples of use can also be found in the Zgoubi Tutorial\cite{tutorial} available online. 
	
\subsection{Handling rotations and translations}

 	     A translation and/or rotation may be required between different elements which can be achieved by using the `CHANGREF' keyword. The CHANGREF keyword is used to define transverse and longitudinal shifts and a z-axis rotation. The entry for this keyword in Part B of the Zgoubi User's Guide is shown in figure \ref{changref} below.
 	     
\begin{figure}[h!]
\centering
\includegraphics[width=\textwidth]{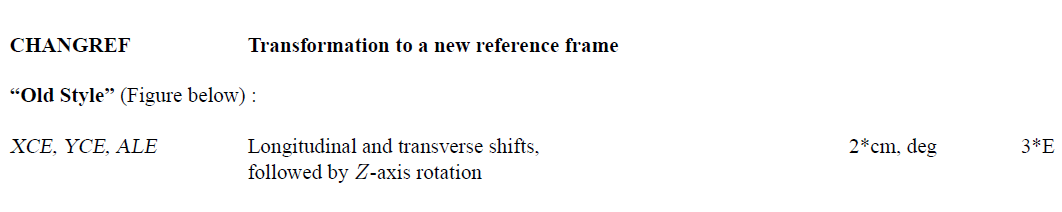}
\caption{An excerpt from the Zgoubi User's Guide \cite{userguide}. }
\label{changref}
\end{figure}

 	         The above entry gives an explanation of the function of the term, the units required (where `2*cm deg' means two separate entries in cm followed by one entry in degrees), and each entry to be in real numbers (as indicated by `3*E'). As such, a rotation of the frame of reference by 10 degrees with respect to the z axis of the original frame would be entered into zgoubi.dat as below.

\begin{lstlisting}
`CHANGREF'
0 0 -10
\end{lstlisting}
\subsection{Defining initial particle coordinates}

	The keyword for the initial particle is 'OBJET'. The input parameters are specified over the subsequent five lines of code as can be seen in the example code below. With reference to Part B of the User's Guide it can be found that the number on the second of these lines (with value '2' in the code below) selects options about the subsequent variables. It is again important to note the units used ($1$\,G = $10^{-4}$\,T). The units for the particle coordinates can be in different units dependent on the number option used on line 2.
	
\begin{lstlisting}
'OBJET' 
-36.689	*Rigidity - note the units of kG.cm : 10.5 eV electron	
2	*Selects menu option 2
1 1	*Total no. of particles, no. of distinct momenta
-0.9 45 0.001 0.01 1  e 	*y, y', z, z', dl, p/p_0
1	*Switch to toggle ray-tracing or not
\end{lstlisting}

	This defined object describes the starting coordinates at the beginning of a cell in the lattice. If these coordinates were all set to zero the particle would travel along the reference path. However this may not be the best starting path, and may result in the particle getting lost after travelling though a few components. Some trial and error is likely to be involved to check if a particle can travel through the lattice correctly. Careful selection of a starting point for a cell - often at a point of symmetry - may mean that the right starting coordinates can be found by varying the horizontal displacement (the Y value) alone. 

\subsection{Extracting local particle coordinates}

	Printing out the local particle coordinates into zgoubi.res can be done with the keyword 'FAISCEAU'. However if the coordinates are to be stored (ie saved to file for later analysis) the keyword 'FAISTORE' should be used. This saves the particle data to a filename (filename.fai) to be specified under the keyword. The keyword 'MARKER' can be used to free-text name a point in the lattice. For example the following code would label a point in the lattice as \textbf{three} and produce a .fai file 'three\_zgoubi.fai' which would record the particle data at every occurrence of the marker 'three'.
	
\begin{lstlisting}
'MARKER'    three
'FAISTORE' 
three_zgoubi.fai three 
2      *This value indicates how frequently the data is recorded, i.e. every second lap around the ring
\end{lstlisting}

\section{Example: The EMMA lattice}

\begin{figure}[h!]
         \centering
         \begin{subfigure}[b]{\textwidth}
                 \centering
                 \includegraphics[width=\textwidth]{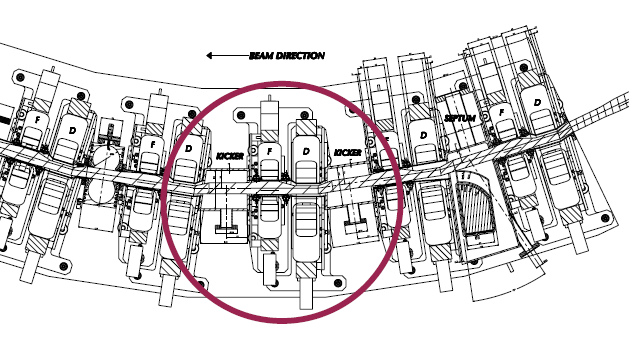}
                 \caption{A schematic drawing of the cell.}
                 \label{fig:schem}
         \end{subfigure}%
         ~ \\
         \begin{subfigure}[b]{\textwidth}
                 \centering
                 \includegraphics[width=\textwidth]{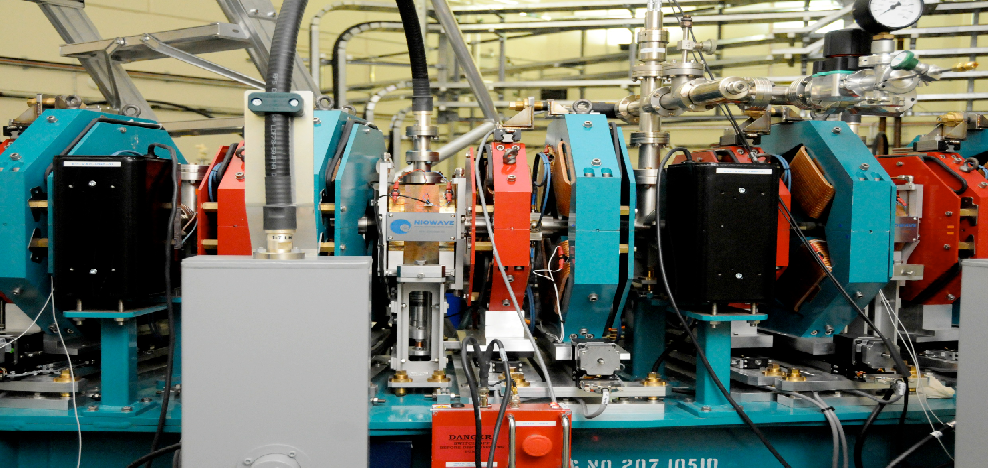}
                 \caption{A photograph of the cell.}
                 \label{fig:pic}
         \end{subfigure}
         ~ \\
         \caption{The EMMA quadrupole cell\cite{mach}. }\label{fig:cell}
\end{figure}

	EMMA (Electron Model of Many Applications) is a non-scaling FFAG accelerator. It consists of 42 doublet quadrupole cells  (see figure \ref{fig:cell}) in a $16.6$\,m circumference ring which accelerate electrons from $10 - 20$\,MeV (kinetic energy)\cite{sue}.

\begin{SCfigure}
\centering
\includegraphics[width=0.4\textwidth]{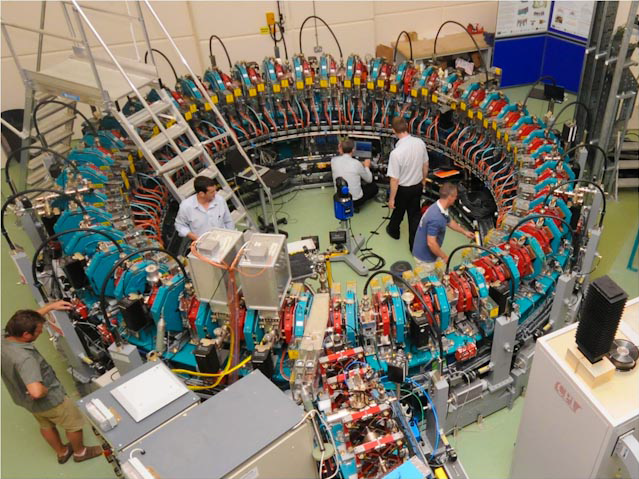}
\caption{The complete EMMA ring\cite{mach}. Each blue and red module is the focusing/defocusing quadrupole doublet, separated from the next doublet by an RF cavity. There are no dipoles used.}\label{fig:ring}
\end{SCfigure}

	As the EMMA ring contains 42 replications of the same doublet cell the lattice definition is comparatively straightforward (see figure \ref{fig:ring}) - after definition the cell can be repeated 42 times to produce the complete ring. 

\subsection{The zgoubi.dat file}

\begin{wrapfigure}{r}{0.5\textwidth}
 
     \includegraphics[width=0.48\textwidth]{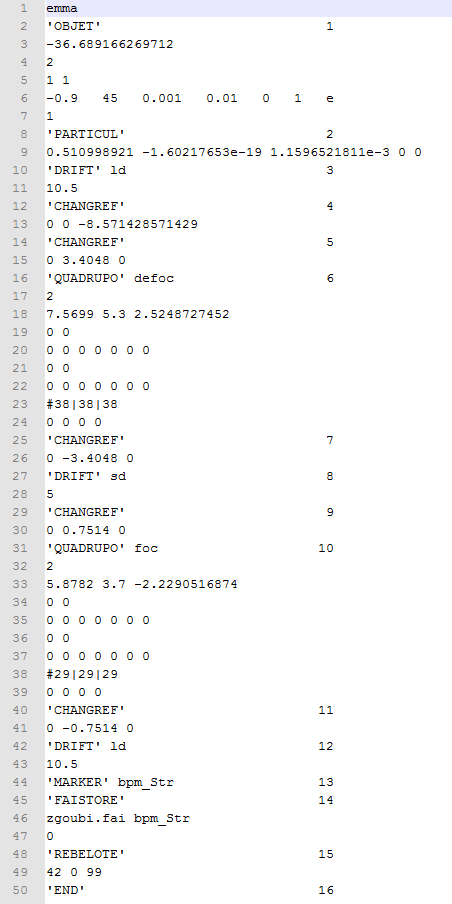}

   \caption{The zgoubi.dat file for the EMMA lattice.}
\label{emmadat}
\end{wrapfigure}

The 'zgoubi.dat' input file for the EMMA lattice can be seen in figure \ref{emmadat}. This section will explain the content of this file line by line using some of the components as described in the previous section.

\begin{figure}
\begin{centering}
\includegraphics[width=\textwidth]{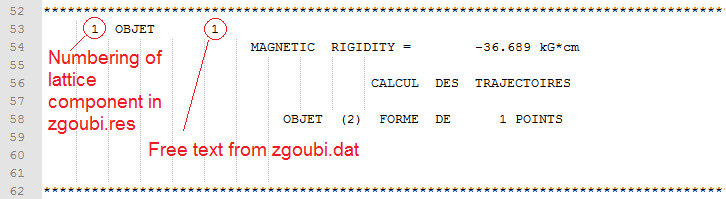}
\caption{Except from zgoubi.res regarding keyword 'OBJET'.}
\label{res1}
\end{centering}
\end{figure}

Along the right hand side of the data in this file it can be seen that each keyword has been labelled in numerical order. This is not an essential part of the file, but can make matching the keyword to the output in the zgoubi.res file more simple. It can be seen in figure \ref{res1} that the results file zgoubi.res numbers each lattice element (the number before the keyword). Labelling each element in the input file with the corresponding number can make it easier to locate an input element with its output. For this lattice, with only 16 keywords used, it is not very difficult to identify each element in the zgoubi.res file and match it to the corresponding element in the zgoubi.dat file but in more complicated lattices this is less straightforward.

\begin{figure}[h!]
\begin{centering}
\includegraphics[width=\textwidth]{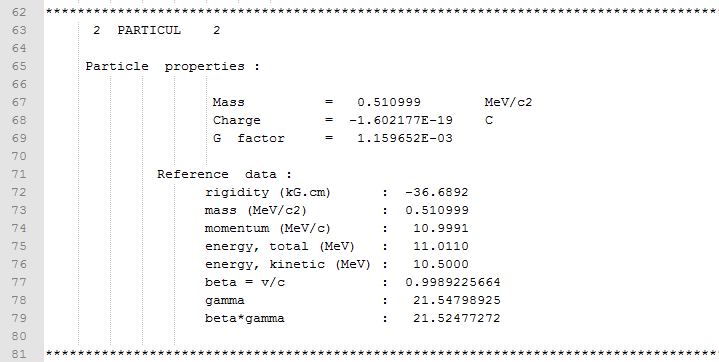}
\caption{Except from zgoubi.res regarding keyword 'PARTICUL'.}
\label{res2}
\end{centering}
\end{figure}

Line 1 of the zgoubi.dat file is for free text, in order to identify the lattice. The first keywords needed for this, and most lattices, are 'OBJET' and 'PARTICUL'. OBJET defines the object, i.e. the initial particle coordinates and PARTICUL defines the characteristics of the particles in the beam. For EMMA the beam consists of electrons.   The zgoubi.res file output gives further details about the particle from the input data from OBJET and PARTICUL (see figure \ref{res2}). It is strongly advised to check this file after defining a lattice to ensure this output is correct, and the data has been input correctly.

\begin{figure}[h!]
\begin{centering}
\includegraphics[width=\textwidth]{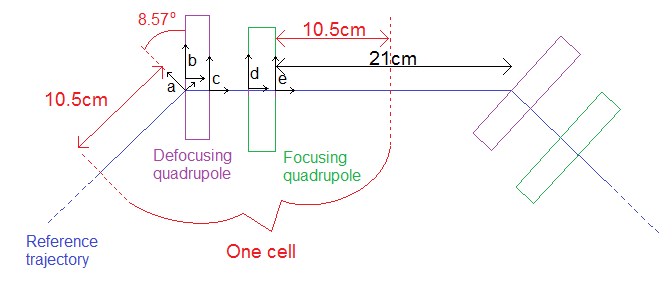}
\caption{Schematic drawing of the cell as defined in zgoubi.dat.}
\label{cell}
\end{centering}
\end{figure}

	At this point in the input file the cell can be defined. In EMMA the cell consists of the quadrupole doublet followed by a drift of $21$\,cm. The next cell is then at an angle to the first. As there are 42 cells in the complete ring this angle is $360^\circ/42 = 8.57^\circ$ (actually $-8.57^\circ$ with respect to the x axis to describe a rotation towards the centre of the ring) . For this lattice the drift section has been described as two $10.5$\,cm sections at the beginning and end of the cell. This means the rotation of reference frame appears within the cell description rather than at the end as expected.  The reason for this relates to the initial injection of particles in the OBJET definition. In order to obtain a stable orbit around the ring some trial and error with the initial particle coordinates is required. Using the middle of the drift space as the start of the cell allows a stable orbit to be found by varying the horizontal displacement (Y) alone. If the start/end of the cell involved the rotation the angle theta (T) would need to be varied too, making the process of finding a stable orbit more complicated. So it can be seen in figure \ref{emmadat} at line 10 the first drift section of $10.5$\,cm is defined, which is then followed by a change in reference frame of a rotation of $-8.57^\circ$. The other half of the drift cavity is defined at the end of the cell on line 42. This defines a cell represented by the schematic picture in figure \ref{cell}. 
	
	The CHANGREF on line 12 of zgoubi.dat rotates the coordinate system from 'a' in figure \ref{cell} to one with  the x-axis parallel to 'b'. The following CHANGREF on line 14 then performs a translation to move the coordinates to 'b', with the x-axis in line with the centre of the defocusing quadrupole. Line 16 then defines the defocusing quadrupole, with the label 'defoc' and is followed by a CHANGREF which reverses the previous translation and moves the coordinates to 'c'. After a further drift section there is another change of coordinates to 'd' before the definition of the focusing quadrupole on line 31. One final CHANGREF moves the coordinates to 'e' before the final drift section is defined at line 42.
	
	The defocusing and focusing quadrupoles are defined at lines 16 and 31 respectively. The initial value of '2' is an instruction to print the field and coordinates along the trajectories. The following lines (18 and 33) define the length, radius and magnetic field at the pole tip for each quadrupole. The focusing magnet can be seen to have a negative value for this field. Both quadrupoles are then defined with four lines consisting of zero values. These lines can be used to define the fringe magnetic fields - the 'entrance' and 'exit' fringe fields. With zero values a 'sharp edge' magnetic field approximation is used. 
	
	On lines 23 and 38 three values are entered to define the 'XPAS' - this defines the number of steps for the integration process in the entrance field, main field and exit field of each component.  If passed a single value a step length in cm would be used.

	At line 44 a marker is placed. This itself has no impact on the beam (if given a second keyword of \textbf{.plt} data from this point in the lattice will be stored in the zgoubi.plt file) but marks the position on the lattice with the label \textbf{bpm\_str}. The following keyword 'FAISTORE' stores particle information in a .fai file given as an argument, here \textbf{zgoubi.fai}. The information is stored at every occurence of the labels given on line 46 which in this case is bpm\_str. So the net result of lines 44-47 is to produce a file output of particle data at the end of the cell where the marker is placed.
	
	The 'REBLOTTE' keyword on line 48 of figure \ref{emmadat} is a command to repeat the zgoubi.dat file. Here it is given the values '42 0 99' which translates as 42 repetitions (i.e. one complete ring) with the '0' specifying a level of verbosity in the \textbf{.res} file and the '99' indicating that the particle coordinates at the end of one pass are used as initial coordinates for the next pass. To track the particle through several laps of the ring a multiple of 42 would be used for the first value.
	
\subsection{Running EMMA and sample output}

\begin{figure}[h!]
	\centering
	\includegraphics[width=0.5\textwidth]{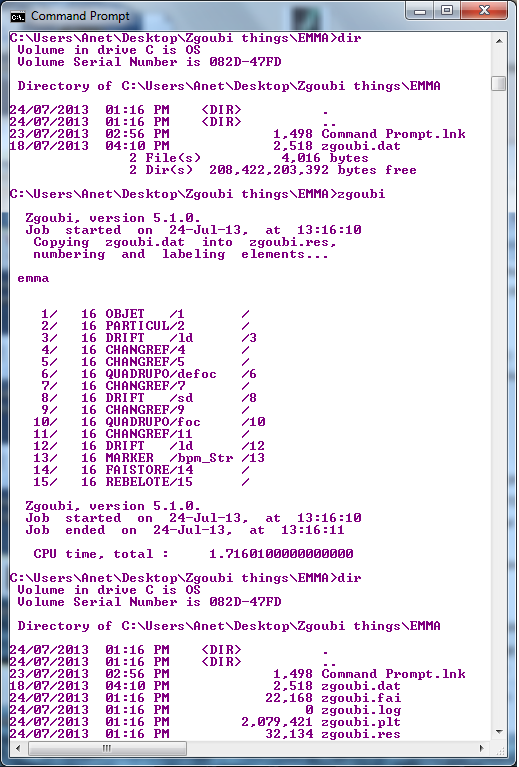}
	\caption{A run of Zgoubi in a Windows command prompt showing the directory contents before and after the run.}
	\label{windowsrun}
	\end{figure}

	A run can be performed by typing the command \textbf{zgoubi} in the directory containing EMMA's zgoubi.dat file (within a command prompt in Windows); this can be seen in figure \ref{windowsrun}, the run produces four output files.
	
\subsubsection{Zpop output}	
\label{sec:zpop}

	If used in UNIX the output can be visualised with Zpop which is part of the Zgoubi package. In the following example the same zgoubi.dat file is used as input as seen in figure \ref{emmadat}. 
	
\begin{figure}
	\vspace{-40pt}
         \centering
         \begin{subfigure}[b]{\textwidth}
                 \centering
                 \includegraphics[width=0.7\textwidth]{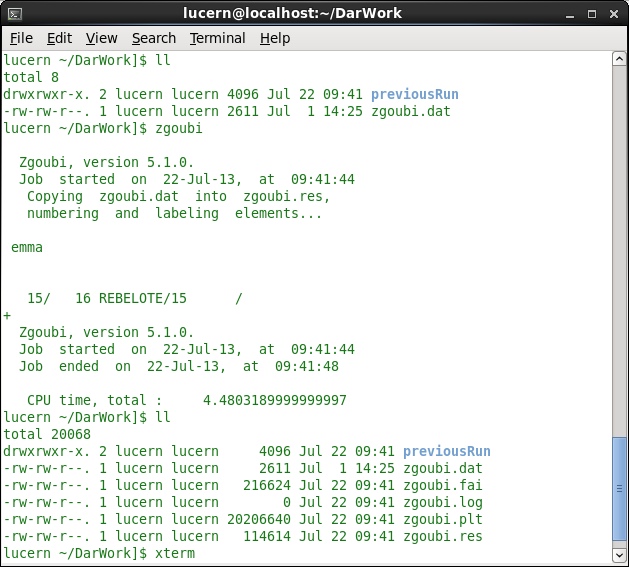}
                 \caption{A run of \textbf{zgoubi.dat}.}
                 \label{unixrun}
         \end{subfigure}%
         ~ \\
         ~\\
          \begin{subfigure}[b]{0.48\textwidth}
                 \centering
                 \includegraphics[width=\textwidth]{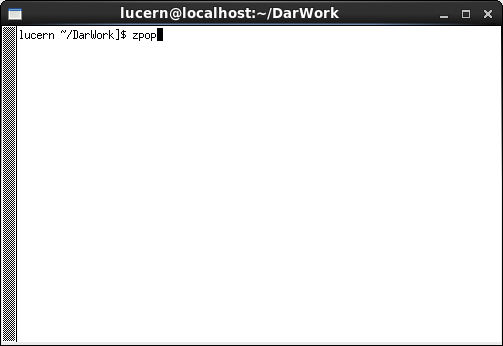}
                \caption{Entering the command \textbf{zpop} in the initial xterm window.}
                 \label{xterm}
         \end{subfigure}
         ~ 
         \begin{subfigure}[b]{0.48\textwidth}
                 \centering
                 \includegraphics[width=\textwidth]{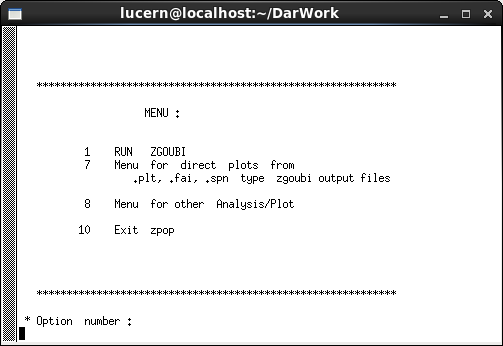}
                 \caption{The initial Zpop menu in the xterm window.}
                 \label{zpop1}
         \end{subfigure}
         ~ \\
         ~\\
         \begin{subfigure}[b]{\textwidth}
                 \centering
                 \includegraphics[width=0.7\textwidth]{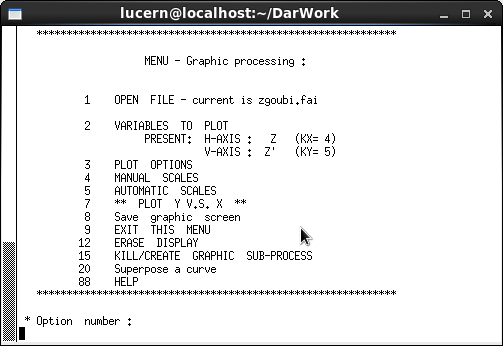}
                 \caption{The menu ready to plot \textbf{Z} vs \textbf{Z'} in the xterm window.}
                 \label{zpop2}
         \end{subfigure}
         \caption{Output from Zgoubi on a Scientific Linux OS.}\label{zgoubiout}
\end{figure}

\begin{figure}
\centering
\includegraphics[width=0.7\textwidth]{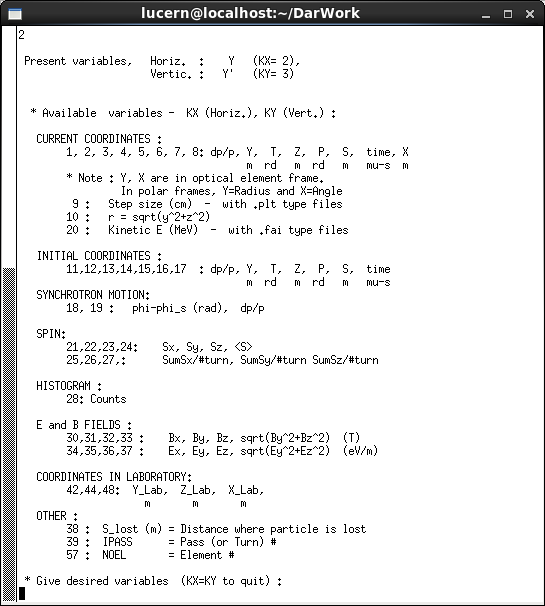}
\caption{The variables available to plot within Zpop for the zgoubi.fai file.}
\label{zpopopt}
\end{figure}

	In order to produce data plots first Zgoubi should be run as in the previous section to produce .fai and/or .plt files, as can be seen in figure \ref{unixrun}. An xterm window can then be opened with the command 'xterm'. In the xterm window the command \textbf{zpop} (see figure \ref{xterm}) then opens the menu \ref{zpop1} and a blank tektronix window. Firstly option 1, 'Run Zgoubi' should be selected in the xterm window to produce a zpop.log file. After this has been done the other options can be selected to examine different data. A list of the available variables to plot can be seen in figure \ref{zpopopt}. There is also an analysis package as part of Zpop accessible through option 8 of the main menu. This offers options as can be seen in figure \ref{zpopa} below.
	
\begin{figure}
\centering
\includegraphics[width=0.7\textwidth]{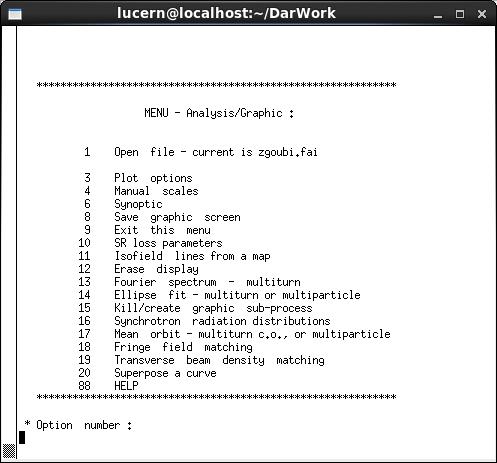}
\caption{The analysis menu in zpop.}
\label{zpopa}
\end{figure}

\begin{figure}
	\centering
	\begin{subfigure}[b]{\textwidth}
	\centering
	\includegraphics[width=0.7\textwidth]{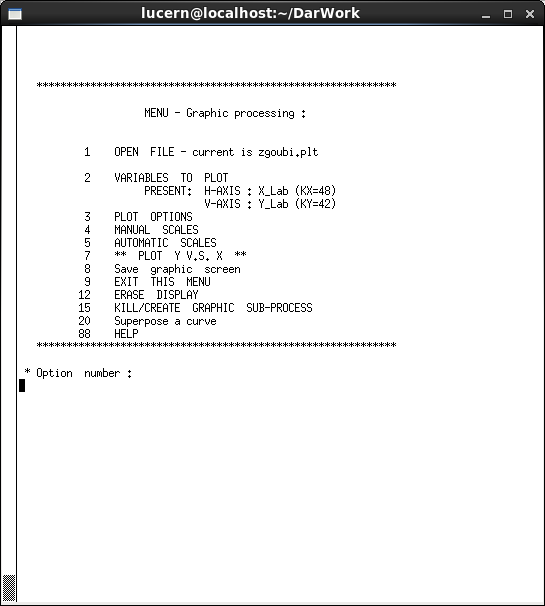}
	\caption{The menu showing the selected options for the plot below using the zgoubi.plt file.}
	\end{subfigure}
	~\\
	\begin{subfigure}[b]{\textwidth}
	\centering
	\includegraphics[width=0.7\textwidth]{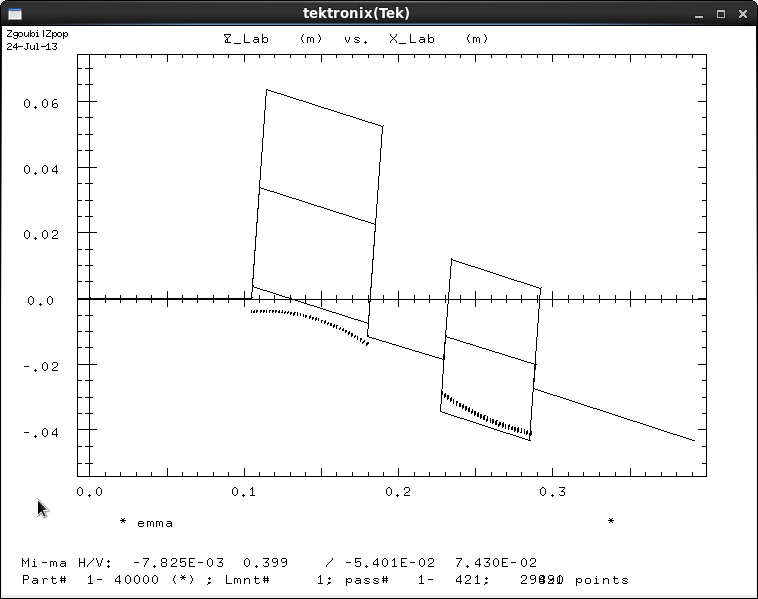}
	\caption{A plot to show the particle track (in lab coordinates) through the magnetic fields of the quadrupoles of one cell. The .plt file only contains tracking data though magnetic fields.}
	\label{pltplot}
	\end{subfigure}
	\caption{An example of using zgoubi.plt in zpop.}
	\label{plt}
	\end{figure}

	As an example of the possible options, it is possible to plot the track of the particle through the lattice quadrupole by loading the zgoubi.plt file (see figure \ref{plt}). Plotting the lab X and Y coordinates then produces a plot showing one cell in the lab coordinates. The quadrupoles can be seen as trapezoidal rather than rectangular shapes as the axis have different scales. The reference track which passes through the centre of the quadrupoles can be seen changing direction in line with the 'CHANGREF' commands from zgoubi.dat. The particle tracking data can be seen as two arcs of data  through the quadrupoles. The trapezoids representing the quadrupoles do not define the physical extent of the magnetic field, so it can be seen that although the particle track happens outside of the defocusing quadrupole trapezoid the track is still curved by its magnetic field. 
	
\begin{figure}
         \centering
         \begin{subfigure}[b]{\textwidth}
                 \centering
                 \includegraphics[width=0.7\textwidth]{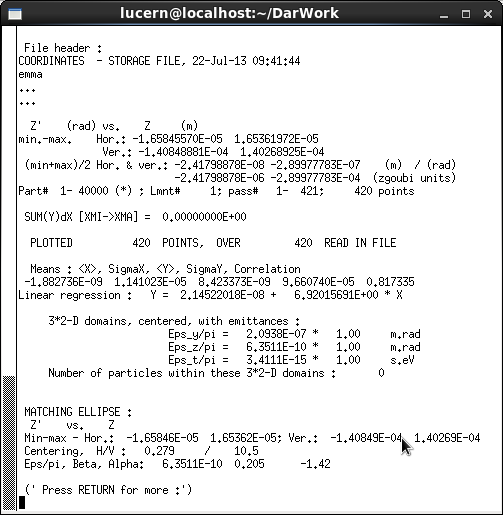}
                 \caption{The xterm output from plotting \textbf{Z} vs \textbf{Z'}.}
                 \label{output}
         \end{subfigure}%
        ~\\
        ~\\
         \begin{subfigure}[b]{\textwidth}
                 \centering
                 \includegraphics[width=0.7\textwidth]{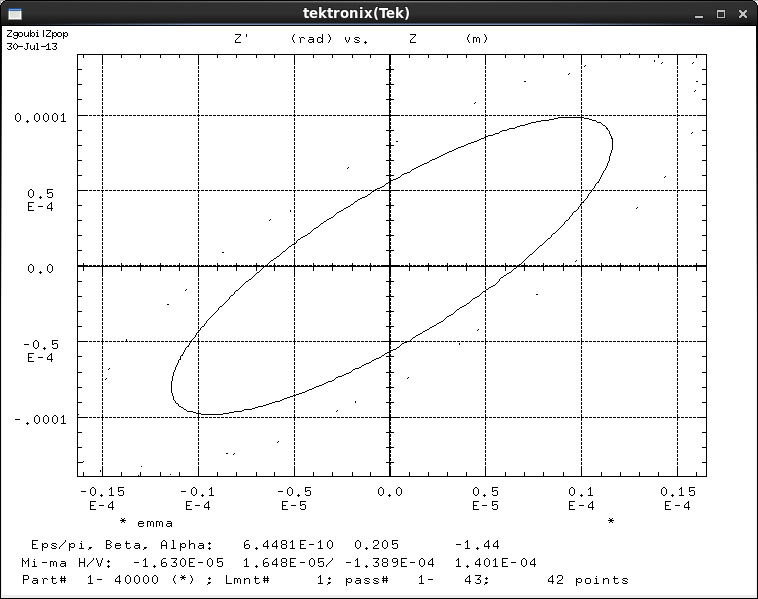}
                 \caption{The graphical output of \textbf{Z} vs \textbf{Z'} in the tektronix window with a 'matched ellipse'.}
                 \label{tek}
         \end{subfigure}
         \caption{Output from Zpop on a Scientific Linux OS.}\label{zpopout}
\end{figure}

	The zgoubi.fai file is useful for analysing the phase space. From the Zpop main menu choosing the menu options to open the file 'zgoubi.fai' allows plotting of a choice of variables. If the variables \textbf{Z} and \textbf{Z'} are chosen (see figure \ref{zpop2}) and then plotted the following phase space chart can be produced (see figure \ref{tek}).

\subsubsection{Analysis in Windows}
	
\begin{figure}
	\vspace{-84pt}
\centering
\begin{subfigure}[b]{\textwidth}
\centering
\includegraphics[width=0.7\textwidth]{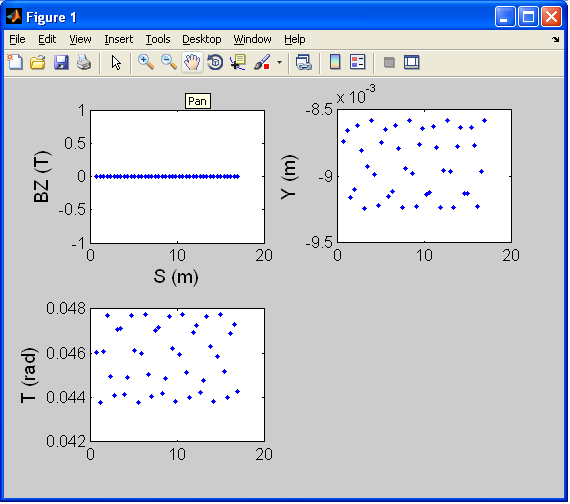}
\end{subfigure}     
~
   \begin{subfigure}[b]{\textwidth}
\centering
\includegraphics[width=0.7\textwidth]{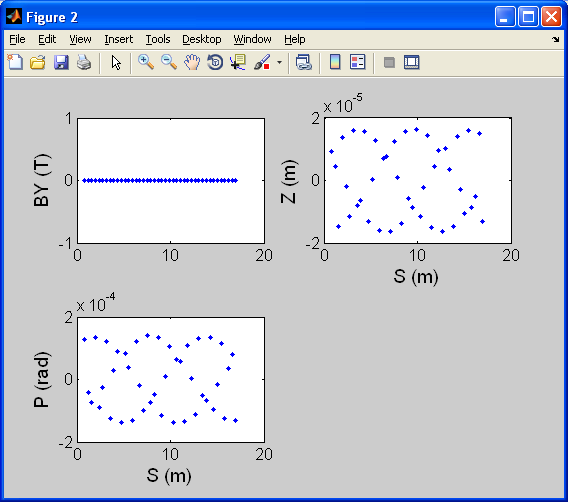}
\end{subfigure}   
~
\begin{subfigure}[b]{\textwidth}
\centering
\includegraphics[width=0.7\textwidth]{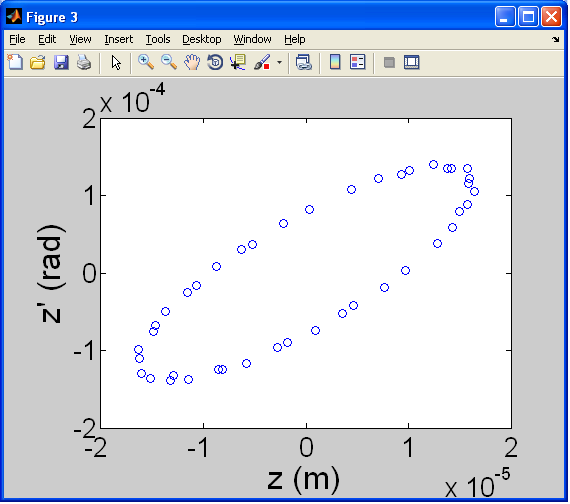}
\end{subfigure}  
\caption{Output from MATLAB scripts analysing zgoubi.fai (see Appendix \ref{sec:scripts}) in Windows.}
\label{matlaboutput}
\end{figure}

	As the running of Zpop requires an xterm window it it not ideally suited to a Windows system. In this situation a MATLAB script can be used to extract the data from the zgoubi.fai or zgoubi.plt file (see Appendix \ref{sec:scripts}). The output from these files can be seen in figure \ref{matlaboutput}.

\subsubsection{EMMA with PyZgoubi}

\begin{figure}[h!]
         \centering
         \begin{subfigure}[b]{\textwidth}
                 \centering
                 \includegraphics[width=0.7\textwidth]{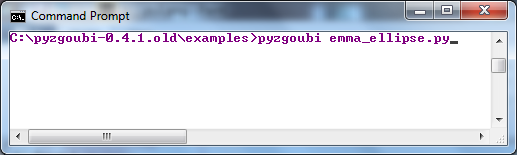}
                 \caption{Entering the command to run.}
                 \label{py1a}
         \end{subfigure}%
        ~\\
        ~\\
         \begin{subfigure}[b]{\textwidth}
                 \centering
                 \includegraphics[width=0.8\textwidth]{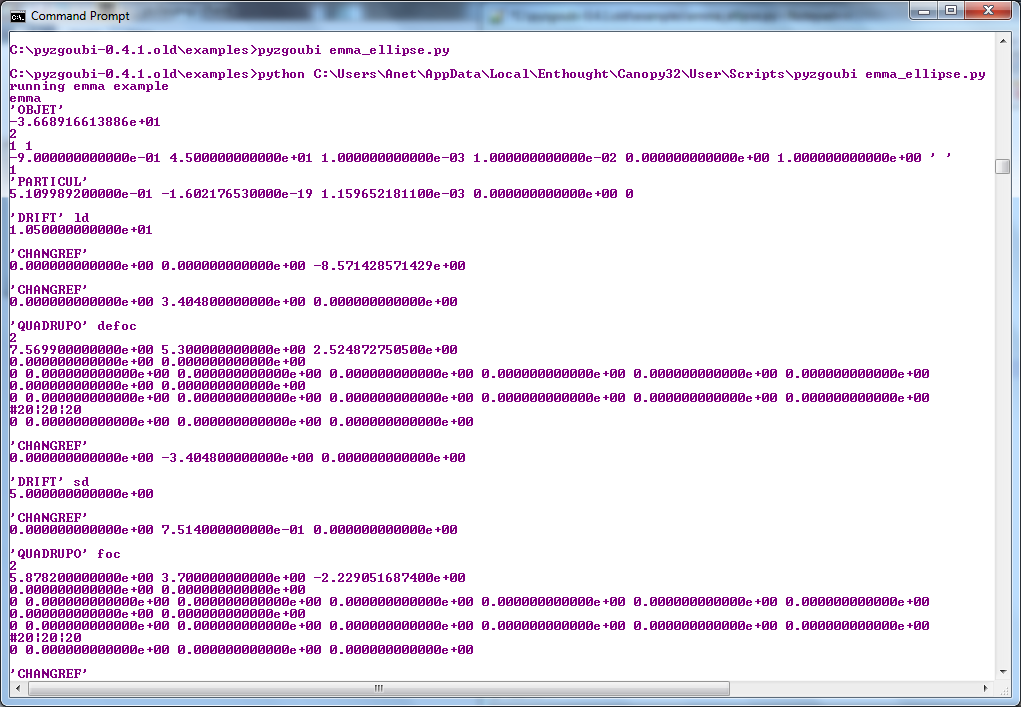}
                 \caption{The initial output of the run.}
                 \label{py1b}
         \end{subfigure}
         ~\\
        ~\\
         \begin{subfigure}[b]{\textwidth}
                 \centering
                 \includegraphics[width=0.7\textwidth]{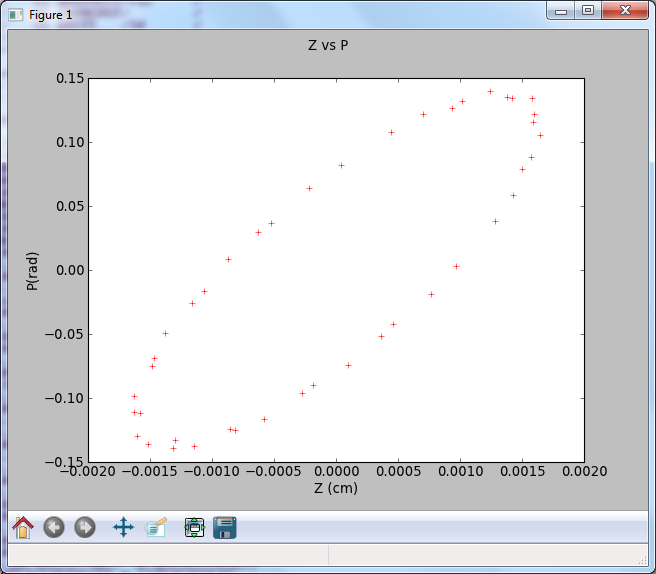}
                 \caption{The graphical output of the script.}
                 \label{py2}
         \end{subfigure}
         \caption{Running \textbf{emma\_ellipse.py} in Windows.}
\end{figure}

	The identical lattice can be defined using the PyZgoubi script found in appendix \ref{sec:emmapy}. A version of this script and other EMMA examples are included in the PyZgoubi downloaded package. As with Zgoubi, the program is run from a command line (cmd.exe in Windows) open in the directory containing the particular script. The 'emma\_ellipse.py' script is run by entering the command \textbf{pyzgoubi emma\_ellipse.py} (see figure \ref{py1a}). The output of the script begins with a print to screen of the zgoubi.res file produced (figure \ref{py1b}), followed by other instructions from the code - in this case a list of the Z,P values and a graphical output (see figure \ref{py2}).
	
	An advantage of PyZgoubi is the ability to use python analysis tools on the resultant data. Included in the PyZgoubi downloaded package are sample EMMA scripts to calculate the closed orbits, tune, Twiss profiles and magnetic apertures for the lattice. Futher information about the use of PyZgoubi can be found at \url{http://www.hep.manchester.ac.uk/u/sam/pyzgoubi/doc/0.3/doc.html}.

\newpage
\section{Appendix}

\subsection{MATLAB scripts}
\label{sec:scripts}
	The following codes were written by Dr. Kai Hock to extract and plot data from the Zgoubi output files using MATLAB. 
\subsubsection{get\_zgoubi\_data.m}	

	This code extracts data from zgoubi.plt.

\begin{lstlisting}
function [Y, T, Z, P, X, S, BX, BY, BZ] = get_zgoubi_data(filename)
% filename = 'zgoubi.plt';
%	 s1 = 'e';    % particle label
	 
fid = fopen (filename, 'r');

fgets(fid);
fgets(fid);
fgets(fid);
fgets(fid);
    
s = fscanf(fid, '\%s', 1);
fgets(fid);

a = fscanf(fid, '\%f', 3);
fgets(fid);
c = fscanf(fid, '\%f', 3);
fgets(fid);
b = fscanf(fid, '\%f', 8);

n = 1;
	
while feof(fid) == 0,
Y(n) = a(2);
T(n) = a(3);
Z(n) = c(1);
P(n) = c(2);
S(n) = c(3);
X(n) = b(5);
BX(n) = b(6);
BY(n) = b(7);
BZ(n) = b(8);
n = n + 1;

fgets(fid);
fgets(fid);
s = fscanf(fid, '\%s', 1);
fgets(fid);

a = fscanf(fid, '\%f', 3);
fgets(fid);
c = fscanf(fid, '\%f', 3);
fgets(fid);
b = fscanf(fid, '\%f', 8);

end

fclose (fid);
\end{lstlisting}

\subsubsection{plt\_zgoubi\_plt\_2b.m}

	In this script the previous code is called to extract the data which is then plotted. 

\begin{lstlisting}
clear

filename = 'zgoubi.plt';
[Y, T, Z, P, X, S, BX, BY, BZ] = get_zgoubi_data(filename);

close all
figure

subplot(221), plot(S/100, BZ/10, '.')
set(gca,'fontsize',12)
xlabel('S (m)', 'fontsize', 14)
ylabel('BZ (T)', 'fontsize', 14)

subplot(222), plot(S/100, Y/100, '.')
set(gca,'fontsize',12)
%xlabel('S (m)', 'fontsize', 14)
ylabel('Y (m)', 'fontsize', 14)

subplot(223), plot(S/100, T/1000, '.')
set(gca,'fontsize',12)
%xlabel('S (m)', 'fontsize', 14)
ylabel('T (rad)', 'fontsize', 14)

figure

subplot(221), plot(S/100, BY/10, '.')
set(gca,'fontsize',12)
%xlabel('S (m)', 'fontsize', 14)
ylabel('BY (T)', 'fontsize', 14)

subplot(222), plot(S/100, Z/100, '.')
set(gca,'fontsize',12)
xlabel('S (m)', 'fontsize', 14)
ylabel('Z (m)', 'fontsize', 14)

subplot(223), plot(S/100, P/1000, '.')
set(gca,'fontsize',12)
xlabel('S (m)', 'fontsize', 14)
ylabel('P (rad)', 'fontsize', 14)

figure

plot(Z/100, P/1000, 'o')
set(gca, 'fontsize', 20)
xlabel('Z (m)', 'fontsize', 24)
ylable('Z'' (rad)', 'fontsize', 24)
\end{lstlisting}
	
	\subsection{PyZgoubi EMMA script}
	\label{sec:emmapy}
	
	This script, emma\_ellipse.py, will write the zgoubi.dat file for the EMMA lattice, print the 'zgoubi.res' output to screen and plot a chart of Z vs Z' (or Z vs P).
	
\begin{lstlisting}
print "running emma example"
import pylab as plt
emma = Line('emma')

xpas = (20,20,20)

cells = 42
angle = -360/cells
d_offset = 34.048 * mm
f_offset = 7.514 * mm

#lengths
ld = 210 * mm
sd = 50 * mm

fq = 58.782 * mm
dq = 75.699 * mm

# quad radius
fr = 37 * mm
dr = 53 * mm

fb = -6.02446402 * fr * T
db = 4.76391085 * dr * T

ob = OBJET2()
emma.add(ob)

emma.add(ELECTRON())

emma.add(DRIFT('ld', XL=ld*cm_/2))
emma.add(CHANGREF(ALE=angle))

emma.add(CHANGREF(YCE=d_offset*cm_))
emma.add(QUADRUPO('defoc', XL=dq*cm_, R_0=dr*cm_, B_0=db*kgauss_, XPAS=xpas, IL=2))
emma.add(CHANGREF(YCE=-d_offset*cm_))

emma.add(DRIFT('sd', XL=sd*cm_))

emma.add(CHANGREF(YCE=f_offset*cm_))
emma.add(QUADRUPO('foc', XL=fq*cm_, R_0=fr*cm_, B_0=fb*kgauss_, XPAS=xpas, IL=2))
emma.add(CHANGREF(YCE=-f_offset*cm_))

emma.add(DRIFT('ld', XL=ld*cm_/2))

emma.add(FAISCNL(FNAME='zgoubi.fai'))

emma.add(REBELOTE(K=99, NPASS=42))

emma.add(END())

rigidity = ke_to_rigidity(10.5e6, 0.51099892e6)
ob.set(BORO=-rigidity)
ob.add(Y=-0.9, T=45, Z=0.001, P=0.01, D=1)

print emma.output()
res = emma.run()#xterm = True)

fai_data =  res.get_all('fai')
print "fai"
print "PASS Z P"
z=[]
q=[]
for p in fai_data:
	z.append(p['Z'])
	q.append(p['P'])
	print p['PASS'], p['Z'], p['P']

fig = plt.figure()
plt.plot(z,q,'r+')
fig.suptitle('Z vs P')
plt.xlabel('Z (m)')
plt.ylabel('P(rad)')
plt.show()	
	
print res.res()
res.clean()

	\end{lstlisting}


\begin{thebibliography}{99}

\bibitem{sourceforge}
"Zgoubi", F. M\'{e}ot, \url{http://sourceforge.net/projects/zgoubi/}

\bibitem{userguide}
"Zgoubi User's Guide", F. M\'{e}ot, 2012, \url{https://www.bnl.gov/isd/documents/80278.pdf}

\bibitem{cern}
"Zgoubi", C. Lapoire, R. Basset, F. Zimmermann and F. Ruggiero, \url{https://oraweb.cern.ch/pls/hhh/code_website.disp_code?code_name=Zgoubi}

\bibitem{examples}
\url{http://www.scienceaccelerator.gov/dsa/result-list/fullRecord:zgoubi/}

\bibitem{matlab}
Mathworks, 2013, \url{http://www.mathworks.co.uk/}

\bibitem{sam}
"PyZgoubi - a python interface to Zgoubi", S.Tygier, \url{http://www.hep.manchester.ac.uk/u/sam/pyzgoubi/}

\bibitem{min}
MinGW, 2013, \url{http://www.mingw.org/}

\bibitem{sam2}
"PyZgoubi Documentation", S. Tygier, \url{http://www.hep.manchester.ac.uk/u/sam/pyzgoubi/doc/0.3/doc.html}


\bibitem{tutorial}
"Zgoubi Tutorial - A course for (potential) users", F. M\'{e}ot, 2009, \url{ftp://ftp3.ie.freebsd.org/pub/sourceforge/z/zg/zgoubi/exemples/tutorial/lecture.pdf}

\bibitem{sue}
"EMMA, The World's First Non-Scaling FFAG Accelerator", S.L. Smith, 2009, \url{http://accelconf.web.cern.ch/AccelConf/PAC2009/papers/we4pbi01.pdf} 

\bibitem{mach}
"Fixed Field Alternating Gradient Accelerators (FFAG)", S.Machida, 2012, \url{http://cas.web.cern.ch/cas/Granada-2012/Lectures/GranadaLectures/Machida.pdf}


\end{thebibliography}
\end{document}